\documentstyle[thmsa,12pt,sw20lart]{article}

\input tcilatex
\QQQ{Language}{
American English
}

\begin{document}

\author{Nick Laskin\thanks{%
E-mail: nlaskin@rocketmail.com, Fax: 1(416) 978 4711}}
\title{{\bf Fractional Schr\"odinger equation}\\
}
\date{University of Toronto\\
IsoTrace Laboratory\\
60 St. George Street, Toronto, ON, M5S 1A7\\
Canada}
\maketitle

\begin{abstract}
Properties of the fractional Schr\"odinger equation have been studied. We
have proven the hermiticity of fractional Hamilton operator and established
the parity conservation law for the fractional quantum mechanics. As
physical applications of the fractional Schr\"odinger equation we have found
the energy spectrum for a hydrogen-like atom - fractional ''Bohr atom'' and
the energy spectrum of fractional oscillator in the semiclassical
approximation. A new equation for the fractional probability current density
has been developed and discussed.

We also discuss the relationships between the fractional and the standard
Schr\"odinger equations.

{\it PACS }number(s): 03.65.-w, 05.30.-d, 05.40. Fb, 03.65. Db

\end{abstract}

\section{Introduction}

The Feynman path integral approach to quantum mechanics \cite{Feynman}, \cite
{Feynman1} is in fact integration over Brownian-like quantum mechanical
paths. The Brownian motion is a special case of  the L\'evy $\alpha $-stable
random process. In the mid of 1930's P. L\'evy and A.Y. Khintchine posed the
question: When the sum of $N$ independent identically distributed quantities 
$X=X_1+X_2...+X_N$ has the same probability distribution $p_N(X)$ (up to
scale factor) as the individual steps $p_i(X_i)$, $i=1,...N$? The
traditional answer is that each $p_i(X_i)$ should be a Gaussian, because of
a central limit theorem. In other words, a sum of $N$ Gaussians is again a
Gaussian. P. L\'evy and A.Y. Khintchine proved that there exists the
possibility to generalize the central limit theorem \cite{Levy}, \cite
{Khintchine}. They discovered class of non-Gaussian L\'evy $\alpha $-stable
(stable under summation) probability distributions. Each $\alpha $-stable
probability distribution has a stability index $\alpha $ often called by the
L\'evy index, $0<\alpha \leq 2$. When $\alpha =2$ the L\'evy $\alpha $%
-stable distribution is transformed to the well-known Gaussian probability
distribution or in other words, the L\'evy motion is transformed to the
Brownian motion \cite{Feller}, \cite{Zolotarev}.

The possibility to develop the path integral over the paths of the L\'evy
motion was discussed by Kac \cite{Kac}, who pointed out that the L\'evy path
integral generates the functional measure in the space of left (or right)
continues functions having only discontinuities of the first kind.

In the Refs.\cite{Laskin1}, \cite{Laskin2} it was shown that the path
integral over the L\'evy-like quantum-mechanical paths allows to develop the
generalization of the quantum mechanics. Namely, if the path integral over
Brownian trajectories leads to the well known Schr\"odinger equation, then
the path integral over L\'evy trajectories leads to the fractional
Schr\"odinger equation. The fractional Schr\"odinger equation includes the
space derivative of order $\alpha $ instead of the second $(\alpha =2)$
order space derivative in the standard Schr\"odinger equation. Thus, the
fractional Schr\"odinger equation is the fractional differential equation in
accordance with the modern terminology (see, for example, \cite{Zaslavsky}-%
\cite{Metzler}). This is the main point for the term fractional
Schr\"odinger equation or for more general term fractional quantum
mechanics, fQM \cite{Laskin2}. As it was mentioned above at $\alpha =2$\ the
L\'evy motion becomes the Brownian motion. Thus, the fQM includes the
standard QM as a particular Gaussian case at $\alpha =2$. Quantum mechanical
path integral over the L\'evy paths at $\alpha =2$ becomes the well known
Feynman path integral \cite{Feynman}, \cite{Feynman1}.

The paper is organized as follows.

In Sec.2 the quantum mechanical path integral over the Levy paths has been
discussed and the 3D fractional Schr\"odinger equation has been derived in
term of the Riesz fractional derivative.

In Sec.3 we prove the hermiticity of fractional Hamilton operator in the
Hilbert space with scalar product defined by the same way as for
conventional quantum mechanics. The parity conservation law for the
fractional quantum mechanics has been established.

Time-independent fractional Schr\"odinger equation has been introduced and
its properties have been studied in Sec.4. As physical applications of the
time-independent fractional Schr\"odinger equation we have found (i) the
energy spectrum and equation for the orbits radius for a hydrogen-like atom
- fractional ''Bohr atom'', (ii) the energy spectrum of 1D fractional
oscillator in the semiclassical approximation.

In Sec.5 a new equation for the fractional probability current density has
been derived and discussed.

In the Conclusion we discuss the relationships between the fractional and
the well-known Schr\"odinger equation.

\section{Path integral}

\subsection{Path integral over the Levy paths}

If a particle at an initial time $t_a$ starts from the point ${\bf r}_a$ and
goes to a final point ${\bf r}_b$ at time $t_b$, we will say simply that the
particle goes from $a$ to $b$ and its path ${\bf r}(t)$ will have the
property that ${\bf r}(t_a)={\bf r}_a$ and ${\bf r}(t_b)={\bf r}_b$. In
quantum mechanics, then, we will have an quantum-mechanical amplitude, often
called a kernel, to get from the point $a$ to the point $b$. This will be
the sum over all of the trajectories that go between that end points and of
a contribution from each. If we have the quantum particle moving in the
potential $V({\bf r})$ then the fractional quantum-mechanical amplitude $K_L(%
{\bf r}_bt_b|{\bf r}_at_a)$ may be written as \cite{Laskin2}

\begin{equation}
K_L({\bf r}_bt_b|{\bf r}_at_a)=  \label{eq1}
\end{equation}

\[
\int\limits_{{\bf r}(t_a)={\bf r}_a}^{{\bf r}(t_b)={\bf r}_b}{\rm D}{\bf r}%
(\tau )\int\limits_{}^{}{\rm D}{\bf p}(\tau )\exp \left\{ \frac i\hbar
\int\limits_{t_a}^{t_b}d\tau [{\bf p}(\tau )\stackrel{\cdot }{\bf r}(\tau
)-H_\alpha ({\bf p}(\tau ),{\bf r}(\tau ))]\right\} , 
\]

where $\hbar $ is the Planck's constant, $\stackrel{\cdot }{\bf r}$ denotes
the time derivative, $H_\alpha ({\bf p}(\tau ),{\bf r}(\tau ))$ is the
fractional Hamiltonian given by 
\begin{equation}
H_\alpha ({\bf p},{\bf r})=D_\alpha |{\bf p}|^\alpha +V({\bf r}),\qquad
1<\alpha \leq 2,  \label{eq2}
\end{equation}

with the replacement ${\bf p}\rightarrow {\bf p}(\tau )$, ${\bf r}%
\rightarrow {\bf r}(\tau )$ and $\{{\bf p}(\tau ),{\bf r}(\tau )\}$ is the
particle trajectory in phase space. The quantity $D_\alpha $ has physical
dimension

\[
\lbrack D_\alpha ]={\rm erg}^{1-\alpha }\cdot {\rm cm}^\alpha \cdot {\rm sec}%
^{-\alpha }. 
\]

The phase space path integral $\int\limits_{{\bf r}(t_a)={\bf r}_a}^{{\bf r}%
(t_b)={\bf r}_b}{\rm D}{\bf r}(\tau )\int\limits_{}^{}{\rm D}{\bf p}(\tau
)...$ in Eq.(\ref{eq1}) is defined by

\begin{equation}
\int\limits_{{\bf r}(t_a)={\bf r}_a}^{{\bf r}(t_b)={\bf r}_b}{\rm D}{\bf r}%
(\tau )\int\limits_{}^{}{\rm D}{\bf p}(\tau )...=  \label{eq3}
\end{equation}

\[
=\stackunder{N\rightarrow \infty }{\lim }\int\limits_{-\infty }^\infty d{\bf %
r}_1...d{\bf r}_{N-1}\frac 1{(2\pi \hbar )^{3N}}\int\limits_{-\infty
}^\infty d{\bf p}_1...d{\bf p}_N\times 
\]

\[
\exp \left\{ i\frac{{\bf p}_1({\bf r}_1-{\bf r}_a)}\hbar -i\frac{D_\alpha
\varsigma |{\bf p}_1|^\alpha }\hbar \right\} \times ...\times \exp \left\{ i%
\frac{{\bf p}_N({\bf r}_b-{\bf r}_{N-1})}\hbar -i\frac{D_\alpha \varsigma |%
{\bf p}_N|^\alpha }\hbar \right\} ..., 
\]

here $\varsigma =(t_b-t_a)/N$.

The exponential in Eq.(\ref{eq1}) can be written as $\exp \{\frac i\hbar
S_\alpha ({\bf p},{\bf r})\}$ if we introduce classical mechanical action
for the trajectory $\{{\bf p}(t)$, ${\bf r}(t)\}$ in phase space

\begin{equation}
S_\alpha ({\bf p},{\bf r})=\int\limits_{t_a}^{t_b}d\tau ({\bf p}(\tau )%
\stackrel{\cdot }{\bf r}(\tau )-H_\alpha ({\bf p}(\tau ),{\bf r}(\tau ),\tau
).  \label{eq4}
\end{equation}

When $\alpha =2$, $D_\alpha =1/2m$, ($m$ is the mass of a particle) Eq.(\ref
{eq2}) is transformed into the well known Hamiltonian with the kinetic
energy ${\bf p}^2/2m$ and Eq.(\ref{eq1}) becomes the definition of the
Feymnam path integral in the phase space representation, see for example 
\cite{Kleinert}.

\subsection{Fractional Schr\"odinger equation}

The kernel $K_L({\bf r}_bt_b|{\bf r}_at_a)$ which is defined by Eq.(\ref{eq1}%
), describes the evolution of the quantum-mechanical system

\begin{equation}
\psi _f({\bf r}_b,t_b)=\int d{\bf r}_aK_L({\bf r}_bt_b|{\bf r}_at_a)\cdot
\psi _i({\bf r}_a,t_a),  \label{eq5}
\end{equation}
where $\psi _i({\bf r}_a,t_a)$ is the wave function of initial (at the $%
t=t_a)$ state and $\psi _f({\bf r}_b,t_b)$ is the wave function of final (at
the $t=t_b)$ state.

In order to obtain the differential equation for the wave function $\psi (%
{\bf r},t)$ we apply Eq.(\ref{eq5}) in the special case that the time
differs only by an infinitesimal interval $\varepsilon $ from $t_a$

\[
\psi ({\bf r},t+\varepsilon )=\int d{\bf r}^{\prime }K_L({\bf r}%
,t+\varepsilon |{\bf r}^{\prime },t)\cdot \psi ({\bf r}^{\prime },t). 
\]

Using Feynman's approximation $\int\limits_t^{t+\varepsilon }d\tau V({\bf r}%
(\tau ),\tau )\simeq \varepsilon V(\frac{{\bf r}+{\bf r}^{\prime }}2,t)$ and
the definition given by Eq.(\ref{eq1}) we have

\[
\psi ({\bf r},t+\varepsilon )= 
\]

\[
\int d{\bf r}^{\prime }\frac 1{(2\pi \hbar )^3}\int\limits_{-\infty }^\infty
d{\bf p}\exp \{i\frac{{\bf p}({\bf r}^{\prime }-{\bf r})}\hbar -i\frac{%
D_\alpha \varepsilon |{\bf p}|^\alpha }\hbar -\frac i\hbar \varepsilon V(%
\frac{{\bf r}+{\bf r}^{\prime }}2,t)\}\cdot \psi ({\bf r}^{\prime },t). 
\]

We may expand the left-hand and the right-hand sides in power series

\begin{equation}
\psi ({\bf r},t)+\varepsilon \frac{\partial \psi ({\bf r},t)}{\partial t}%
=\int d{\bf r}^{\prime }\frac 1{(2\pi \hbar )^3}\int\limits_{-\infty
}^\infty d{\bf p}\exp \{i\frac{{\bf p}({\bf r}^{\prime }-{\bf r})}\hbar
\}\cdot (1-i\frac{D_\alpha \varepsilon |{\bf p}|^\alpha }\hbar )\times
\label{eq6}
\end{equation}

\[
(1-\frac i\hbar \varepsilon V(\frac{{\bf r}+{\bf r}^{\prime }}2,t))\cdot
\psi ({\bf r}^{\prime },t). 
\]

Then, taking into account the definitions of the Fourier transforms

\[
\psi ({\bf r},t)=\frac 1{(2\pi \hbar )^3}\int d{\bf p}e^{i\frac{px}\hbar
}\varphi ({\bf p},t),\qquad \varphi ({\bf p},t)=\int d{\bf p}e^{-i\frac{{\bf %
px}}\hbar }\psi ({\bf r},t), 
\]

and introducing the 3D quantum Riesz fractional derivative\footnote{%
The Riesz fractional derivative was originally introduced in \cite{Riesz}} $%
(-\hbar ^2\Delta )^{\alpha /2}$

\begin{equation}
(-\hbar ^2\Delta )^{\alpha /2}\psi ({\bf r},t)=\frac 1{(2\pi \hbar )^3}\int
d^3pe^{i\frac{{\bf pr}}\hbar }|{\bf p}|^\alpha \varphi ({\bf p},t),
\label{eq7}
\end{equation}

(here $\Delta =\partial ^2/\partial {\bf r}^2$ is the Laplacian) we obtain
from Eq.(\ref{eq6})

\[
\psi ({\bf r},t)+\varepsilon \frac{\partial \psi ({\bf r},t)}{\partial t}%
=\psi ({\bf r},t)-i\frac{D_\alpha \varepsilon }\hbar (-\hbar ^2\Delta
)^{\alpha /2}\psi ({\bf r},t)-\frac i\hbar \varepsilon V({\bf r},t)\psi (%
{\bf r},t). 
\]

This will be true to order $\varepsilon $ if $\psi ({\bf r},t)$ satisfies
the differential equation

\begin{equation}
i\hbar \frac{\partial \psi ({\bf r},t)}{\partial t}=D_\alpha (-\hbar
^2\Delta )^{\alpha /2}\psi ({\bf r},t)+V({\bf r},t)\psi ({\bf r},t).
\label{eq8}
\end{equation}

This is the fractional Schr\"odinger equation. The space derivative in this
equation is of fractional (noninteger) order $\alpha $.

The above consideration is in fact the generalization of the well known
Feynman approach to reduce the path integral to the differential equation 
\cite{Feynman}, \cite{Feynman1}.

Equation (\ref{eq8}) may be rewritten in the operator form, namely

\begin{equation}
i\hbar \frac{\partial \psi }{\partial t}=H_\alpha \psi ,  \label{eq9}
\end{equation}

where $H_\alpha $ is the fractional Hamiltonian operator

\begin{equation}
H_\alpha =D_\alpha (-\hbar ^2\Delta )^{\alpha /2}+V({\bf r},t).  \label{eq10}
\end{equation}

By definition (\ref{eq10}) and introducing the momentum operator ${\bf p}%
=i\hbar {\bf \nabla }$ one may obtain the fractional Hamiltonian $H_\alpha $
in the form given by Eq.(\ref{eq2}).

Since the kernel $K_L({\bf r}_bt_b|{\bf r}_at_a)$ thought of as a function
of variables ${\bf r}_b,t_b$, is a special wave function (namely, that for a
particle which starts at ${\bf r}_a,t_a$), we see that $K_L$ must also
satisfy a fractional Schr\"odinger equation. Thus for the quantum system
described by the fractional Hamiltonian Eq.(\ref{eq10}) we have

\begin{equation}
i\hbar \frac \partial {\partial t_b}K_L({\bf r}_bt_b|{\bf r}_at_a)=D_\alpha
(-\hbar ^2\Delta _b)^{\alpha /2}K_L({\bf r}_bt_b|{\bf r}_at_a)+V({\bf r}%
_b,t)K_L({\bf r}_bt_b|{\bf r}_at_a),  \label{eq11}
\end{equation}

where $t_b>t_a,$ and the low index ''$b$'' at the $\Delta _b$ means that the
fractional derivative acts on the variable ${\bf r}_b$.

\section{Quantum Riesz fractional derivative}

\subsection{Hermiticity of the fractional Hamilton operator}

The fractional Hamiltonian $H_\alpha $ given by Eq.(\ref{eq10}) is the
Hermitian operator in the space with scalar product

\begin{equation}
(\phi ,\chi )=\int\limits_{-\infty }^\infty d{\bf r}\phi ^{*}({\bf r},t)\chi
({\bf r},t),  \label{eq12}
\end{equation}

where the sign $*$ means as usual complex conjugate.

To prove hermiticity of the fractional Hamilton $H_\alpha $ let us note that
in accordance with definition of the quantum Riesz fractional derivative
given by Eq.(\ref{eq7}) there exists the integration by parts formula

\begin{equation}
(\phi ,(-\hbar ^2\Delta )^{\alpha /2}\chi )=((-\hbar ^2\Delta )^{\alpha
/2}\phi ,\chi ).  \label{eq13}
\end{equation}

The average energy of fractional quantum system with Hamiltonian $H_\alpha $
is

\begin{equation}
E_\alpha =\int\limits_{-\infty }^\infty d{\bf r}\psi ^{*}({\bf r},t)H_\alpha
\psi ({\bf r},t).  \label{eq14}
\end{equation}

Taking into account Eq.(\ref{eq13}) we have

\[
E_\alpha =\int\limits_{-\infty }^\infty d{\bf r}\psi ^{*}({\bf r},t)H_\alpha
\psi ({\bf r},t)=\int\limits_{-\infty }^\infty d{\bf r}(H_\alpha ^{+}\psi (%
{\bf r},t))^{*}\psi ({\bf r},t)=E_\alpha ^{*}. 
\]

and as a physical consequence, the energy of a system is real. Thus, the
fractional Hamiltonian $H_\alpha $ defined by Eq.(\ref{eq10}) is the
Hermitian or self-adjoint operator in the space with the scalar product
defined by Eq.(\ref{eq12})

\begin{equation}
(H_\alpha ^{+}\phi ,\chi )=(\phi ,H_\alpha \chi ).  \label{eq15}
\end{equation}

\subsection{Parity conservation law for the fractional quantum mechanics}

It follows from the definition (\ref{eq7}) of the quantum Riesz fractional
derivative that

\begin{equation}
(-\hbar ^2\Delta )^{\alpha /2}\exp \{i\frac{{\bf px}}\hbar \}=|{\bf p}%
|^\alpha \exp \{i\frac{{\bf px}}\hbar \}.  \label{eq16}
\end{equation}

Thus, the function $\exp \{i{\bf px}/\hbar \}$ is the eigenfunction of the
3D quantum Riesz fractional operator $(-\hbar ^2\Delta )^{\alpha /2}$ with
eigenvalue $|{\bf p}|^\alpha $.

The operator $(-\hbar ^2\Delta )^{\alpha /2}$ is symmetrized fractional
derivative, that is

\begin{equation}
(-\hbar ^2\Delta _{{\bf r}})^{\alpha /2}...=(-\hbar ^2\Delta _{-{\bf r}%
})^{\alpha /2}....  \label{eq17}
\end{equation}

Because of the property (\ref{eq17}) the fractional Hamiltonian $H_\alpha $
given by Eq.(\ref{eq10}) remains invariant under {\it inversion}
transformation. Inversion, or to be precise, spatial inversion consists in
the simultaneous change in sign of all three spatial coordinates

\begin{equation}
{\bf r}\rightarrow -{\bf r},\qquad x\rightarrow -x,\quad y\rightarrow
-y,\quad z\rightarrow -z.  \label{eq18}
\end{equation}

Let us denote the inversion operator by $\widehat{P}$. The inverse symmetry
is the fact that $\widehat{P}$ and the fractional Hamiltonian $H_\alpha $
commute,

\begin{equation}
\widehat{P}H_\alpha =H_\alpha \widehat{P}.  \label{eq19}
\end{equation}

We can divide the wave functions of quantum mechanical states with a
well-defined eigenvalue of the operator $\widehat{P}$ into two classes; (i)
functions which are not changed when acted upon by the inversion operator,

\[
\widehat{P}\psi _{+}({\bf r})=\psi _{+}({\bf r}); 
\]

the corresponding states are called even states; (ii) functions which change
sign under the action of the inversion operator,

\[
\widehat{P}\psi _{-}({\bf r})=-\psi _{-}({\bf r}); 
\]

the corresponding states are called odd states.

Equation (\ref{eq19}) express the ''parity conservation law''; if the state
of a closed fractional quantum mechanical system has a given parity (i.e. if
it is even, or odd), then this parity is conserved.

\section{Time-independent fractional Schr\"odinger equation}

The special case when the Hamiltonian $H_\alpha $ does not depend explicitly
on the time is of great importance for physical applications. It is easily
to see that in this case there exist the special solution of the fractional
Schr\"odinger equation (\ref{eq8}) of the form

\begin{equation}
\psi ({\bf r},t)=e^{-(i/\hbar )Et}\phi ({\bf r}),  \label{eq20}
\end{equation}

where $\phi ({\bf r})$ satisfies

\begin{equation}
H_\alpha \phi ({\bf r})=E\phi ({\bf r}),  \label{eq21}
\end{equation}

or

\begin{equation}
D_\alpha (-\hbar ^2\Delta )^{\alpha /2}\phi ({\bf r})+V({\bf r})\phi ({\bf r}%
)=E\phi ({\bf r}),\qquad 1<\alpha \leq 2.  \label{eq22}
\end{equation}

The equation (\ref{eq22}) we call by the time-independent (or stationary)
fractional Schr\"odinger equation.

\subsection{Fractional Bohr atom}

The hydrogenlike potential energy $V({\bf r})$ is

\[
V({\bf r})=-\frac{Ze^2}{|{\bf r}|}. 
\]

Then the fractional Schr\"odinger equation (\ref{eq22}) has a form

\begin{equation}
D_\alpha (-\hbar ^2\Delta )^{\alpha /2}\phi ({\bf r})-\frac{Ze^2}{|{\bf r|}}%
\phi ({\bf r})=E\phi ({\bf r}),  \label{eq23}
\end{equation}

and can be treated as fractional eigenvalue problem.

The total energy of considered quantum mechanical system is

\[
E=E_{kin}+V, 
\]

where $E_{kin}$ is the kinetic energy

\begin{equation}
E_{kin}=D_\alpha |{\bf p}|^\alpha ,  \label{eq24}
\end{equation}

and $V$ is the potential energy

\begin{equation}
V=-\frac{Ze^2}{|{\bf r|}}.  \label{eq25}
\end{equation}

It is well known that if the potential energy is a homogeneous function of
the co-ordinates and the motion takes place in a finite region of space,
there exists a simple relation between the time average values of the
kinetic and potential energies, known as the {\it virial theorem }(see, page
23, \cite{Landau1}). It follows from the virial theorem that between average
kinetic energy (\ref{eq24}) and average potential energy (\ref{eq25}) there
exist the relation

\begin{equation}
\alpha \overline{E}_{kin}=-\overline{V},  \label{eq26}
\end{equation}

where the average value $\overline{f}$ of any function of time is defined as

\[
\overline{f}=\stackunder{T\rightarrow \infty }{\lim }\frac
1T\int\limits_0^\infty dtf(t). 
\]

In order to evaluate the energy spectrum of the fractional hydrogenlike atom
let us remind the {\it Niels Bohr postulates} \cite{Bohr}:

1. The electron moves in orbits restricted by the requirement that the
angular momentum be an integral multiple of $\hbar $, that is, for circular
orbits of radius $a_n$, the electron momentum is restricted by

\begin{equation}
pa_n=n\hbar ,\qquad (n=1,2,3,...),  \label{eq27}
\end{equation}

and furthermore the electrons in these orbits do not radiate in spite of
their acceleration. They were said to be in stationary states.

2. Electrons can make discontinuous transitions from one allowed orbit
corresponding to $n=n_2$ to another corresponding to $n=n_1$, and the change
in energy will appear as radiation with frequency

\begin{equation}
\omega =\frac{E_{n_2}-E_{n_1}}\hbar ,\qquad (n_2>n_1).  \label{eq28}
\end{equation}

An atom may absorb radiation by having its electrons make a transition to a
higher energy orbit.

Using the first Bohr's postulate and Eq.(\ref{eq26}) yields

\[
\alpha D_\alpha \left( \frac{n\hbar }{a_n}\right) ^\alpha =\frac{Ze^2}{a_n}, 
\]

from which it follows the equation for the radius of the fractional Bohr
orbits

\begin{equation}
a_n=a_0n^{\frac \alpha {\alpha -1}},  \label{eq29}
\end{equation}

here $a_0$ is the fractional Bohr radius (the radius of the lowest, $n=1$
Bohr orbit) defined as,

\begin{equation}
a_0=\left( \frac{\alpha D_\alpha \hbar ^\alpha }{Ze^2}\right) ^{\frac
1{\alpha -1}}.  \label{eq30}
\end{equation}

By using Eq.(\ref{eq26}) we find for the total average energy $\overline{E}$

\[
\overline{E}=(1-\alpha )\overline{E}_{kin}. 
\]

Thus, for the energy levels of the fractional hydrogen-like atom we have

\begin{equation}
E_n=-(\alpha -1)E_0n^{-\frac \alpha {\alpha -1}},\qquad 1<\alpha \leq 2,
\label{eq31}
\end{equation}

where $E_0$ is the binding energy of the electron in the lowest Bohr orbit,
that is, the energy required to put it in a state with $E=0$ corresponding
to $n=\infty $,

\begin{equation}
E_0=\left( \frac{(Ze^2)^\alpha }{\alpha ^\alpha D_\alpha \hbar ^\alpha }%
\right) ^{\frac 1{\alpha -1}}.  \label{eq32}
\end{equation}

The energy $(\alpha -1)E_0$ can be considered as generalization of the
Rydberg constant of the standard quantum mechanics. It is easy to see that
at $\alpha =2$ the energy $(\alpha -1)E_0$ is transformed into the well
known expression for the Rydberg constant, ${\rm Ry}=me^4/2\hbar ^2$.

The frequency of the radiation $\omega $ associated with the transition,
say, for example from $k$ to $n$, $k\rightarrow n$, is,

\begin{equation}
\omega =\frac{(\alpha -1)E_0}\hbar \cdot \left[ \frac 1{n^{\frac \alpha
{\alpha -1}}}-\frac 1{k^{\frac \alpha {\alpha -1}}}\right] ,\qquad (k>n)
\label{eq33}
\end{equation}

The new equations (\ref{eq29})-(\ref{eq33}) give generalization of the
''Bohr atom'' theory. In a special Gaussian case, $\alpha =2$ (standard
quantum mechanics) Eqs.(\ref{eq29})-(\ref{eq33}) reproduce the well known
results of the Bohr theory \cite{Bohr}, \cite{Bohr1}. The existence of Eqs.(%
\ref{eq29})-(\ref{eq33}) is a result of deviation of fractal dimension ${\rm %
d}_{{\rm fractal}}^{(L\acute evy)}$ of the L\'evy-like quantum mechanical
path from 2, ${\rm d}_{{\rm fractal}}^{(L\acute evy)}=\alpha <2$.

\subsection{Spectrum of the 1D fractional oscillator in semiclassical
approximation}

Fractional oscillator introduced in \cite{Laskin1} is the model with the
fractional Hamiltonian operator $H_{\alpha ,\beta }$,

\begin{equation}
H_{\alpha ,\beta }=D_\alpha (-\hbar ^2\Delta )^{\alpha /2}+q^2|{\bf r}%
|^\beta ,\quad 1<\alpha \leq 2,\quad 1<\beta \leq 2,  \label{eq34}
\end{equation}

where ${\bf r}$ is the 3D vector, $\Delta =\partial ^2/\partial {\bf r}^2$
is the Laplacian, the operator $(-\hbar ^2\Delta )^{\alpha /2}$ is defined
by the Eq.(\ref{eq7}) and $q$ is a constant with physical dimension $[q]=%
{\rm erg}^{1/2}\cdot {\rm cm}^{-\beta /2}$.

The 1D fractional oscillator with the Hamilton function $H_{\alpha ,\beta
}=D_\alpha |p|^\alpha +q^2|x|^\beta $ poses an interesting problem for
semiclassical treatment. We set the total energy equal to $E$, so that

\begin{equation}
E=D_\alpha |p|^\alpha +q^2|x|^\beta ,  \label{eq35}
\end{equation}

whence

\[
|p|=\left( \frac 1{D_\alpha }(E-q^2|x|^\beta )\right) ^{1/\alpha }. 
\]

At the turning points $p=0$. Thus, classical motion is possible in the range 
$|x|\leq (E/q^2)^{1/\beta }$.

A routine use of the Bohr-Sommerfeld quantization rule \cite{Landau} yields

\begin{equation}
2\pi \hbar (n+\frac 12)=\oint pdx=4\int\limits_0^{x_m}pdx=\frac 4{D_\alpha
^{1/\alpha }}\int\limits_0^{x_m}(E-q^2|x|^\beta )^{1/\alpha }dx,
\label{eq36}
\end{equation}

where the notation $\oint $ means the integral over one complete period of
the classical motion, $x_m=(E/q^2)^{1/\beta }$ is the turning point of
classical motion. To evaluate the integral in the right hand of Eq.(\ref
{eq36}) we introduce a new variable $y=x(E/q^2)^{-1/\beta }$. Then we have

\begin{equation}
\int\limits_0^{x_m}(E-q^2|x|^\beta )^{1/\alpha }dx=\frac 1{q^{2/\beta
}}E^{\frac 1\alpha +\frac 1\beta }\int\limits_0^1dy(1-y^\beta )^{1/\alpha }.
\label{eq37}
\end{equation}

The integral over $dy$ can be expressed in the terms of the $B$-function%
\footnote{%
The $B(a,b)$ function has the familiar integral representation \cite{Erdelyi}
\par
$B(a,b)=\int\limits_0^1duu^{a-1}(1-u)^{b-1}$.}. Indeed, substitution $%
z=y^\beta $ yields

\begin{equation}
\int\limits_0^1dy(1-y^\beta )^{1/\alpha }=\frac 1\beta
\int\limits_0^1dzz^{\frac 1\beta -1}(1-z)^{\frac 1\alpha }=\frac 1\beta
B(\frac 1\beta ,\frac 1\alpha +1).  \label{eq38}
\end{equation}

With help of Eqs.(\ref{eq37}) and (\ref{eq38}) we rewrite Eq.(\ref{eq36}) as

\[
2\pi \hbar (n+\frac 12)=\frac 4{D_\alpha ^{1/\alpha }q^{2/\beta }}E^{\frac
1\alpha +\frac 1\beta }\frac 1\beta B(\frac 1\beta ,\frac 1\alpha +1). 
\]

The above equation gives the value of the energies of stationary states for
1D fractional oscillator,

\begin{equation}
E_n=\left( \frac{\pi \hbar \beta D_\alpha ^{1/\alpha }q^{2/\beta }}{2B(\frac
1\beta ,\frac 1\alpha +1)}\right) ^{\frac{\alpha \beta }{\alpha +\beta }%
}\cdot (n+\frac 12)^{\frac{\alpha \beta }{\alpha +\beta }}.  \label{eq39}
\end{equation}

This new equation generalize the well known energy spectrum of the standard
quantum mechanical oscillator (see for example, \cite{Landau}) and is
transformed to it at the special case $\alpha =2$, $\beta =2$.

As it follows from Eq.(\ref{eq39}) at

\begin{equation}
\frac{\alpha \beta }{\alpha +\beta }=1  \label{eq40}
\end{equation}
the energy spectrum becomes equidistant. When $1<\alpha \leq 2$ and $1<\beta
\leq 2$ the condition given by Eq.(\ref{eq40}) takes place for $\alpha =2$
and $\beta =2$ only. It means that only standard quantum mechanical
oscillator has the equidistant energy spectrum.

\section{Current density}

By multiplying Eq.(\ref{eq8}) from left by $\psi ^{*}({\bf r},t)$ and the
conjugate complex of Eq.(\ref{eq8}) by $\psi ({\bf r},t)$ and subtracting
the two resultant equations we finally obtain

\begin{equation}
\frac \partial {\partial t}\int d^3r\left( \psi ^{*}({\bf r},t)\psi ({\bf r}%
,t)\right) =  \label{eq41}
\end{equation}

\[
\frac{D_\alpha }{i\hbar }\int d^3r\left( \psi ^{*}({\bf r},t)(-\hbar
^2\Delta )^{\alpha /2}\psi ({\bf r},t)-\psi ({\bf r},t)(-\hbar ^2\Delta
)^{\alpha /2}\psi ^{*}({\bf r},t)\right) . 
\]

From this integral relationship we are led to the following well known
differential equation

\begin{equation}
\frac{\partial \rho ({\bf r},t)}{\partial t}+{\rm div}{\bf j}({\bf r},t)=0,
\label{eq42}
\end{equation}

where

\begin{equation}
\rho ({\bf r},t)=\psi ^{*}({\bf r},t)\psi ({\bf r},t),  \label{eq43}
\end{equation}

is the density of probability and the vector ${\bf j}({\bf r},t)$ can be
called by the fractional probability current density vector

\begin{equation}
{\bf j}({\bf r},t)=\frac{D_\alpha \hbar }i\left( \psi ^{*}({\bf r},t)(-\hbar
^2\Delta )^{\alpha /2-1}{\bf \nabla }\psi ({\bf r},t)-\psi ({\bf r}%
,t)(-\hbar ^2\Delta )^{\alpha /2-1}{\bf \nabla }\psi ^{*}({\bf r},t)\right) ,
\label{eq44}
\end{equation}

where we use the following notation

\[
{\bf \nabla =}\frac \partial {\partial {\bf r}}. 
\]

Introducing the momentum operator $\widehat{{\bf p}}=\frac \hbar i{\bf %
\nabla }$ we can write the vector ${\bf j}$ in the form

\begin{equation}
{\bf j=}D_\alpha \left( \psi (\widehat{{\bf p}}^2)^{\alpha /2-1}\widehat{%
{\bf p}}\psi ^{*}+\psi ^{*}(\widehat{{\bf p}}^{*2})^{\alpha /2-1}\widehat{%
{\bf p}}^{*}\psi \right) ,\qquad 1<\alpha \leq 2.  \label{eq45}
\end{equation}

When $\alpha =2$, $D_\alpha =1/2m$ Eqs.(\ref{eq44}), (\ref{eq45})\ becomes
the well known equations of the standard quantum mechanics (see, for example 
\cite{Landau}). Thus we conclude that the new Eqs.(\ref{eq44}) and (\ref
{eq45}) are the fractional generalization of the well known equations for
probability current density vector of standard quantum mechanics.

To this end, we express Eq.(\ref{eq45}) in the terms of the velocity
operator, which is defined as usual

\[
\widehat{{\bf v}}=\frac d{dt}\widehat{{\bf r}}, 
\]

where $\widehat{{\bf r}}$ is the operator of coordinate. Using the general
quantum mechanical rule for differentiation of operator

\[
\frac d{dt}\widehat{{\bf r}}=\frac i\hbar [H_\alpha ,{\bf r}], 
\]

we have

\[
\widehat{{\bf v}}=\frac i\hbar (H_\alpha {\bf r-r}H_\alpha ). 
\]

Further, with help of the equation

\[
{\rm f}(\widehat{{\bf p}}){\bf r}-{\bf r}{\rm f}(\widehat{{\bf p}})=-i\hbar 
\frac{\partial {\rm f}}{\partial {\bf p}}, 
\]

which holds for any function ${\rm f}(\widehat{{\bf p}})$ of the momentum
operator, and taking into account Eq.(\ref{eq2}) for the Hamiltonian we find
the equation for the velocity operator

\begin{equation}
\widehat{{\bf v}}=\alpha D_\alpha |\widehat{{\bf p}}^2|^{\alpha /2-1}%
\widehat{{\bf p}},  \label{eq46}
\end{equation}

here $\widehat{{\bf p}}$ is the momentum operator. By comparing of Eqs.(\ref
{eq45}) and (\ref{eq46}) we finally conclude that

\begin{equation}
{\bf j=}\frac 1\alpha \left( \psi \widehat{{\bf v}}\psi ^{*}+\psi ^{*}%
\widehat{{\bf v}}\psi \right) ,\qquad 1<\alpha \leq 2.  \label{eq47}
\end{equation}

To get the probability current density equal 1 (the current when one
particle pass through the unit area per unit time) the wave function of a
free particle have to be normalized as follows

\begin{equation}
\psi ({\bf r},t)=\sqrt{\frac \alpha {2{\rm v}}}\exp \{\frac i\hbar {\bf pr}%
-\frac i\hbar Et\},\qquad E=D_\alpha |{\bf p}|^\alpha ,\qquad 1<\alpha \leq
2,  \label{eq48}
\end{equation}

where ${\rm v}$ is the particle velocity, ${\rm v}=\alpha D_\alpha p^{\alpha
-1}$. Indeed, by substituting Eq.(\ref{eq48}) into Eq.(\ref{eq45}) we find

\begin{equation}
{\bf j=}\frac{{\bf v}}{{\rm v}},\qquad {\bf v}=\alpha D_\alpha |{\bf p}%
^2|^{\frac \alpha 2-1}{\bf p,}  \label{eq49}
\end{equation}

that is the vector ${\bf j}$ is the unit vector.

Equations (\ref{eq44}), (\ref{eq45}) and (\ref{eq47}) are the fractional
generalization of the well known equations for probability current density
vector of the standard quantum mechanics \cite{Landau}.

\section{Conclusions}

Fractional generalization of the Schr\"odinger equation has been studied. We
have established the integration by parts formula for the quantum Riesz
fractional derivative and used it to prove hermiticity of the fractional
Hamilton operator. The parity conservation law for the fractional quantum
mechanics has been observed. The time-independent fractional Schr\"odinger
equation has been introduced. As physical applications of the
time-independent fractional Schr\"odinger equation we have found the energy
spectrum and equation for the orbits radius for a hydrogen-like atom -
fractional ''Bohr atom''. The energy spectrum of 1D fractional oscillator
has been obtained in the semiclassical approximation.

The generalization of the fractional probability current density has been
derived and discussed.

The generalized equations (\ref{eq8}), (\ref{eq22}), (\ref{eq29})-(\ref{eq33}%
), (\ref{eq39}), (\ref{eq44}), (\ref{eq45}) and (\ref{eq47}) are transformed
into the well known equations of conventional quantum mechanics if we put
the L\'evy index $\alpha =2$. In other words the fractional quantum
mechanics includes the standard quantum mechanics as a particular Gaussian
case at $\alpha =2$. Quantum mechanical path integral over the L\'evy paths
at $\alpha =2$ becomes the well known Feynman path integral and the
fractional Schr\"odinger equation becomes the Schr\"odinger equation.

The fractional Schr\"odinger equation provides us with general point of view
on the relationship between statistical properties of quantum-mechanical
path and structure of the fundamental equations of quantum mechanics.

\end{document}